\documentclass[sigconf, natbib]{acmart}
\usepackage[utf8]{inputenc}
\usepackage{enumitem}

\usepackage{multirow} 
\usepackage{makecell} 
\usepackage{subcaption}
\usepackage{textcomp}
\usepackage{tabularx} 
\usepackage{xspace}

\newcommand{\paratitle}[1]{\vspace{1.5ex}\noindent\textbf{#1}}
\newcommand{\method}{DFGR\xspace}

\setlist[itemize]{noitemsep, topsep=-10pt}
\AtBeginDocument{%
  }

\copyrightyear{2025}
\acmYear{2025}
\setcopyright{acmlicensed}\acmConference[CIKM '25]{Proceedings of the 34th ACM International Conference on Information and Knowledge Management}{November 10--14, 2025}{Seoul, Republic of Korea}
\acmBooktitle{Proceedings of the 34th ACM International Conference on Information and Knowledge Management (CIKM '25), November 10--14, 2025, Seoul, Republic of Korea}
\acmDOI{10.1145/3746252.3761106}
\acmISBN{979-8-4007-2040-6/2025/11}

\begin{document}

\title{Action is All You Need: Dual-Flow Generative Ranking Network for Recommendation}


\author{Hao Guo}
\authornote{Corresponding Author.}
\email{guohao15@meituan.com}
\orcid{0000-0002-8955-5146}
\affiliation{%
  \institution{Meituan}
  \city{BeiJing}
  \country{China}
}

\author{Erpeng Xue}
\email{xueerpeng@meituan.com}
\affiliation{%
  \institution{Meituan}
  \city{BeiJing}
  \country{China}
}

\author{Lei Huang}
\email{huanglei45@meituan.com}
\affiliation{%
  \institution{Meituan}
  \city{BeiJing}
  \country{China}
}

\author{Shihao Wang}
\email{wangshichao10@meituan.com}
\affiliation{%
  \institution{Meituan}
  \city{BeiJing}
  \country{China}
}

\author{Xiaolei Wang}
\authornote{This work was done during the internship at Meituan.}
\email{xiaoleiwang@ruc.edu.cn}
\affiliation{%
  \institution{Gaoling School of Artificial Intelligence, Renmin University of China}
  \city{BeiJing}
  \country{China}
}

\author{Zeshun Li}
\authornotemark[2]
\email{lzs23@mails.tsinghua.edu.cn}
\affiliation{%
  \institution{Tsinghua University}
  \city{BeiJing}
  \country{China}
}

\author{Lei Wang}
\email{wanglei46@meituan.com}
\affiliation{%
  \institution{Meituan}
  \city{BeiJing}
  \country{China}
}

\author{Jinpeng Wang}
\email{wangjinpeng04@meituan.com}
\affiliation{%
  \institution{Meituan}
  \city{BeiJing}
  \country{China}
}

\author{Sheng Chen}
\email{chensheng19@meituan.com}
\affiliation{%
  \institution{Meituan}
  \city{BeiJing}
  \country{China}
}

\renewcommand{\shortauthors}{Hao Guo et al.}

\begin{abstract}

 Deep Learning Recommendation Models (DLRMs) often rely on extensive manual feature engineering to improve accuracy and user experience, which increases system complexity and limits scalability of model performance with respect to computational resources. Recently, Meta introduced a generative ranking paradigm based on HSTU block that enables end-to-end learning from raw user behavior sequences and demonstrates scaling law on large datasets that can be regarded as the state-of-the-art (SOTA). However, splitting user behaviors into interleaved item and action information significantly increases the input sequence length, which adversely affects both training and inference efficiency. To address this issue, we propose the Dual-Flow Generative Ranking Network (DFGR), that employs a dual-flow mechanism to optimize interaction modeling, ensuring efficient training and inference through end-to-end token processing. DFGR duplicates the original user behavior sequence into a real flow and a fake flow based on the authenticity of the action information, and then defines a novel interaction method between the real flow and the fake flow within the QKV module of the self-attention mechanism. This design reduces computational overhead and improves both training efficiency and inference performance compared to Meta’s HSTU-based model. Experiments on both open-source and real industrial datasets show that DFGR outperforms DLRM, which serves as the industrial online baseline with extensive feature engineering, as well as Meta’s HSTU and other common recommendation models such as DIN, DCN, DIEN, and DeepFM. Furthermore, we investigate optimal parameter allocation strategies under computational constraints, establishing DFGR as an efficient and effective next-generation generative ranking paradigm.

 
\end{abstract}



\maketitle
\section{Introduction}
In modern recommendation systems, the estimation of the Click-Through Rate (CTR) and the Click-Through Conversion Rate (CTCVR) constitutes a fundamental but critical task. These predictive metrics serve directly as the basis for the distribution of services or products within the system. Any optimization of CTR and CTCVR can be directly translated into enhanced traffic value for the platform. Over the past decade in industry applications, state-of-the-art approaches in recommendations have relied on Deep Learning Recommendation Models (DLRMs). These models leverage multilayer neural networks with a custom-designed architecture to effectively model user interests~\cite{zhang2022dhen,wang2021dcn,xia2023transact}.  However, these models rely heavily on extensive and manually crafted feature engineering, where domain-specific features are meticulously designed by algorithm engineers based on raw system logs. From an anthropocentric perspective, such human-engineered features act as systematic probes. From a macro-perspective, such human-engineered features can be viewed as slice-based probing results of a certain dimension of the system. Regardless of their granularity, they inherently suffer from information loss during input representation, which ultimately limits the performance ceiling of the recommendation system. 

Meta recently proposed a new architecture called HSTU \cite{zhai2024actions} for generative recommendation. HSTU eliminates the need for complex manual feature engineering required by traditional DLRM paradigms, instead utilizing raw user behavior sequences from log systems that comprehensively capture both positive and negative interactions. Each token in the sequence only requires an identification feature and minimal business attributes to ensure generalizability. In the generative ranking module, items and their corresponding action types are organized in an interleaved pattern to better model the interactions between candidate items and sequence elements. The forward process follows the same decoder-only architecture as ChatGPT~\cite{ouyang2022training} and LLaMA~\cite{grattafiori2024llama,touvron2023llama}. 
During training, when computing the loss in backpropagation, the outputs at action-type positions will be masked to enable end-to-end training. This approach evidently uses two tokens to represent each complete user interaction, thereby doubling the input sequence length. Within the self-attention framework, this expansion leads to quadratic growth in computational and memory requirements, imposing substantial overhead that presents significant deployment challenges for industrial applications with strict latency constraints during inference. Furthermore, decomposing each interaction into two semantically distinct tokens creates a heterogeneous sequence where the tokens exhibit a fundamentally unbalanced information density, introducing considerable learning difficulties for the model. For convenience in the subsequent explanation of this paper, we refer to Meta's HSTU-based generative recommendation architecture as MetaGR.

\begin{table}[t]
  \caption{Comparison among MetaGR, and our proposed generative ranking networks, including single-flow~(SFGR) and Dual-flow~(DFGR). Assuming all models employ the same Transformer hyperparameter configurations and input settings. $N$ denotes the number of items in the sequence, while $K$ represents the average number of items per session.}
  \label{tab:intro}
  \begin{tabular}{l|ccc}
    \toprule
    \textbf{Model} & {\begin{tabular}[c]{@{}c@{}}\textbf{Training}\\\textbf{Complexity}\end{tabular}} & {\begin{tabular}[c]{@{}c@{}}\textbf{Inference}\\\textbf{Complexity}\end{tabular}} \\
    \midrule
    MetaGR & 4O($N^2$) & 4O($N^2$) \\
    \midrule
    SFGR & $\frac{1}{3}$O($\frac{N^3}{K}$) & O($N^2$)\\
    DFGR & 2O($N^2$) & O($N^2$) \\
    \bottomrule
  \end{tabular}
\end{table}

We propose an intuitive forward-inference efficiency optimization strategy for the MetaGR baseline approach. This strategy called Single-Flow Generate Ranking network (SFGR) merges item information with action-type in the user's historical sequence, while masking the action-type for items in the candidate set. Specifically, we represent each complete user interaction with a single token, enabling a user sequence of length N to be represented by a token sequence of the same length. During the prediction phase, a user sequence of length N is transformed into a homogeneous token input sequence of length N, reducing the input length of token sequence by approximately 50\% compared to the MetaGR approach, thus significantly improving the prediction inference efficiency. However, this design presents challenges for training efficiency. Assuming that the user behavior sequence can be viewed as $S$ session sequences ordered by timestamp, each session is extracted as training samples while prepending its corresponding preceding user sequences to the training sequence shown in Figure \ref{fig:train_ineffect}. This processing method results in a single complete user sequence being divided into multiple training samples, and only calculating loss for candidate elements after a forward computation which leads to computational resource waste. The fundamental cause of this issue lies in the dual representation roles of items in the user sequence, where they serve both as samples and features. This dual identity directly determines which action-types corresponding to the sequence need to be masked.

To maximize the efficiency of generative ranking networks across training and inference stages while enabling more sophisticated end-to-end feature engineering, we propose a Dual-Flow Generative Ranking Network (DFGR) architecture. When processing the raw user input sequence, it is duplicated into two sequence data flows based on whether the action-type uses the real action type: one flow uses the real action-type identifiers called the real flow, and the other uses fake action-type identifiers, referred to as the fake flow. These two data flows share network parameter information, which allows the action-type token information to be merged as side information into the corresponding item token within each sequence flow, ensuring that each interaction corresponds to only a single token in the sequence. During the training phase, the real flow propagates forward using standard self-attention mechanisms, whereas the fake flow self-attention uses keys and values (KV) derived from the hidden state embeddings of the corresponding layers in the real flow. This ensures that each token in the fake flow perceives the historical sequence fully without revealing the action-type. The loss computation via backpropagation is exclusively based on the output of the fake flow. During the inference stage, we implement the same efficient method described in SFGR, which concatenates the candidate item sequence with the user's historical behavior sequence, enabling parallel scoring through a single forward pass. The training and inference complexity comparison among MetaGR, SFGR and DFGR is presented in Table \ref{tab:intro} which will be proved in Section 3.

Building upon the DFGR architecture, we investigate the scaling law phenomenon specific to the ranking module within the recommendation scenario. We explore optimal parameter design schemes for the model under computational constraints, conducting comprehensive investigations into key components and strategies such as full domain sequence data and learning rate scheduling strategies. These efforts provide crucial guidance for subsequent iterative work.

Overall, we make the following contributions:

\textbullet~We propose a Dual-Flow Generative Ranking Network, namely \textbf{\method}, which enables end-to-end modeling directly from raw user behavioral sequences without requiring heavy feature engineering, while significantly reducing computational costs during both training and inference stages compared to the current state-of-the-art MetaGR paradigm.

\textbullet~We demonstrated the effectiveness of the common recommendation system optimization techniques, such as the mixture of attention MOA\cite{zhang2022mixture}, full-domain features within this architecture in real industrial data, and systematically investigated the optimal parameter allocation strategy under computational constraints.

\textbullet~In both open-source and real industrial datasets, our approach achieved the best offline AUC/G-AUC metrics compared to competitive DLRM baselines that utilize extensive feature engineering and MetaGR approaches, positioning it as the next-generation state-of-the-art generative ranking paradigm.

\section{Related work}
Our work is related to the following research directions.


\paratitle{User Behavior Sequence Modeling.}
The user behavior sequence modeling encompasses a wealth of information that reflects the interest and preferences of the user. In previous modeling approaches based on the DLRM paradigm, significant advances have been achieved in user behavior sequences \cite{cao2022sampling,si2024dual,chang2023dual,pi2019practice,pi2020search,zhou2019deep,feng2019deep,wei2021eta,chen2022efficient}. Using longer sequence lengths and incorporating a greater diversity of behavioral types within the sequences always results in a better metric that indicates that the sequence contains a wealth of information that can effectively characterize user interests. A more comprehensive and in-depth representation of the sequence can often enhance the extraction of user interests.

\paratitle{Feature Interaction Modeling.}
Feature interaction modeling is crucial in DLRM as it captures complex interactions between features\cite{wang2021dcn,guo2017deepfm,shan2016deep,wen2019pairwise,huang2019fibinet,liu2019feature,cheng2016wide,he2017neural,xiao2017attentional}, improving the accuracy of prediction. A strategic approach to feature interaction modeling allows systems to detect subtle patterns and correlations, yielding recommendations that are better aligned with user preferences and behaviors, which is highly effective for improving recommendation metrics and significantly enhancing user experience. Moreover, feature interaction modeling enables dynamic adaptation to changing user behaviors, which is essential to achieve state-of-the-art performance in modern recommendation systems.


\paratitle{Generative Recommendation.}
Meta~\cite{zhai2024actions} recently introduced a generative recommendation model paradigm centered around a decoder-only Transformer architecture, achieving structural unification of the recall and ranking modules at the model architecture level. Within the ranking module, to better capture interactions between candidate items and user sequence elements, item and action information are interleaved during sequence construction. By implementing a strategic mask in the output layer, the network facilitates end-to-end training. During inference, multiple candidate sets are flattened and appended to the sequence, optimizing computational efficiency for parallel predictions. This approach yielded significant benefits in Meta's core short video recommendation scenario.

\section{Methodology}
In this section, we present our proposed dual-flow generative ranking network \textbf{\method}, which demonstrates remarkable improvements in the efficiency for both training and inference.

\subsection{Overview of the Approach}
\paratitle{Task Formulation.}
In recommendation systems, let $\mathcal{U}$ denote a set of users $u$, and $\mathcal{V}$ be a set of items $v$, the ranking tasks (\textit{e.g.} CTR prediction, CTCVR prediction) can be represented as:
\begin{equation}
 Y=f\left(X_{u,v}; \Theta_f\right)
\end{equation} 
where \(f\) is the model function with learnable model parameters \(\Theta_f\), and \(X_{u,v}\) represents a set of characteristics associated with the user and the item. In the DLRM paradigm, \(X_{u,v}\) is mainly constructed by hand engineering of features, combining statistical features of different dimensions that can be regarded as tabular data. 

\begin{figure}[t]
    \centering
    \includegraphics[width=1.0\linewidth]{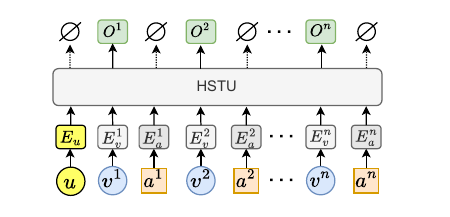}
    \caption{MetaGR's Generative Ranking Architecture}
    \label{fig:hstu}
\end{figure}

\paratitle{Generative Ranking.}
In the generating ranking paradigm, \(X_{u,v}\) is organized in the historical behavior sequence of the user, \textit{i.e.}, \(\{v^1,v^2,\cdots,v^n\}\), where \(n\) denotes the total number of interactions in the sequence. Specifically, each item \(v\) is associated with some side information (e.g., category, geohash, week, hour, city, price), which fully represents when (timestamp), where (geolocation), what (item category) and how (action type) of user interactions.
MetaGR \cite{zhai2024actions} proposed an interleaved way of organizing samples for generative ranking, allowing the ranking model to be trained in an autoregressive manner. For any user \(u \in \mathcal{U} \), the model sample is organized as follows:
\begin{equation}
 S_u=(u, v^1,a^1,\cdots,v^i,a^i,\cdots,v^n,a^n)
\end{equation} 
where  \(u\) denotes the user profile features, \(v^i\) denotes the \(i\)-th interacted item in the sequence, \(a^i \in \{0,1\}\) denotes the action type of the \(i\)-th interacted item. For example, in the CTR prediction task, \(a^i=1\) signifies a click, while \(a^i=0\) indicates no click. The way of organizing samples that intermingles item tokens and action type tokens leads to a doubling of the input sequence length. This significantly increases the computational load for model training and inference, potentially hindering its implementation in real-world industrial settings. As shown in Figure \ref{fig:hstu}, during training, the output corresponding to the action-type is masked, and only the item part of the output is used to calculate the loss. During inference, the action-type information of the candidate part is masked and then concatenated to the user's historical behavior sequence for parallel scoring.
 
\paratitle{Dual Flow For Efficient Generative Ranking.}
Although generative ranking enables autoregressive training, the interleaved way of sample organizing introduces extra training and inference overhead.
To address this, we first observe that items and corresponding action types in user interaction history can be merged for inference and propose a single-flow generative ranking network (Section~\ref{sec:single-flow}).
To further decrease the overhead of training, we disentangle user history sequence and candidate item sequence into two independent flows for encoding, which is the dual-flow generative ranking network (Section~\ref{sec:dual-flow}).

\subsection{Efficient Generative Ranking via Single-Flow Inference}
\label{sec:single-flow}

\begin{figure*}[t]
    \centering
    \includegraphics[width=0.85\textwidth]{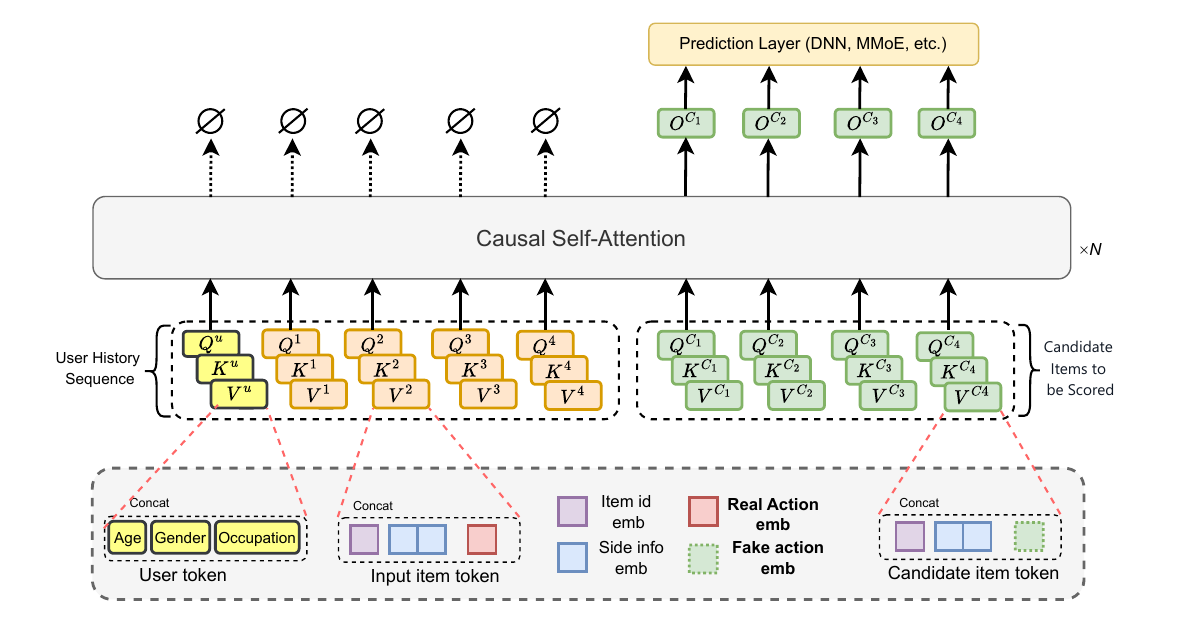}
    \caption{Model Architecture of Single-Flow Generative Ranking Network.}
    \label{fig:sfgr}
\end{figure*}

\begin{figure}[t]
    \centering
    \includegraphics[width=1.0\linewidth]{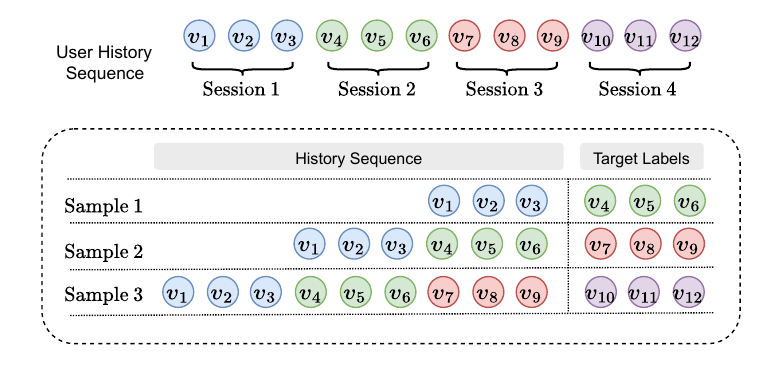}
    \caption{Training Sample Construction Method under the Single-Flow Generative Ranking Network Architecture.}
    \label{fig:train_ineffect}
\end{figure}

\subsubsection{Model Architecture}
To address the high computational cost caused by item element decomposition in MetaGR's basic approach, we propose a single-flow generative ranking network~(SFGR) that merges items and action types for more efficient inference, as shown in Figure ~\ref{fig:sfgr}. Firstly, following the method shown in Figure ~\ref{fig:train_ineffect}, we segment each complete user behavior sequence into multiple subsequences based on session boundaries. As illustrated, the input comprises two components: the user profile with historical behavior sequence (represented by the yellow and orange segments) constructed according to Section 3.1's specifications with preserved actual actions, and candidate items for scoring (represented by the green segment) where real action information is masked using placeholder tokens. This concatenated sequence is processed through multiple layers of a decoder-only Transformer-based network, where, during training, the output elements corresponding to the historical behavior sequence are masked, while the output embeddings of candidate sequence elements are passed through various prediction heads such as MMOE~\cite{ma2018modeling} or other architectures to calculate cross-entropy loss.

At inference time,  we partition all candidates into microbatches of size m to enable parallel scoring across microbatches, where the m candidates $C_1,C_2,....C_m$ within each batch are sequentially flattened in any order while maintaining consistent spatiotemporal context features. To ensure that the ordering of candidate items does not affect their scoring results, we additionally zero out the attention scores between candidate sequence items on the standard lower triangular attention matrix. In this way, all scores for $C_1,C_2,...C_m$ can be obtained in parallel after a single forward pass. 

Furthermore, owing to the decoder-only architecture, these candidates can maximally leverage the computational power of the user profile and user history sequence parts, providing substantial advantages for the model's online serving deployment. 

\subsubsection{Comparative Analysis of Computational Complexity}
Assuming  the model use HSTU as the fundamental computational unit and receive the same user raw sequence input of length $N$, with a hidden dimension of $D$ in HSTU, training batch size $B$, head number $H$, each head corresponds to dimensions $d_q/d_k/d_v/d_u$ sized $d$, and it is assumed that $D=H*d$, layer num is $L$. Excluding parameters related to the embedding part, most of the computation and intermediate activation parameters are concentrated within the network dominated by HSTU modules, thus allowing for a direct comparison of computational focusing on the HSTU units.

\paratitle{Infernece.} 
As for MetaGR, the computational complexity is $2*O(4*B*2N*D^2 + 2*B*H*(2N)^2*d + B*2N*D^2) * L$. 
As for SFGR, the computational complexity is $2*O(4*B*N*D^2 + 2*B*H*N^2*d + B*N*D^2)*L$. 
By using an approximate calculation, we can obtain that SFGR achieves a 4× speedup over the MetaGR model.

\paratitle{Training.} 
As for SFGR, since a user sample is divided into multiple sequences, assuming the average length of each session in the sequence is $K$, then one sequence can be divided into $N/K$ samples. Therefore, the total computational complexity is $\sum_{i=1}^{N/K}2*O(4*B*i*K*D^2 + 2*B*H*(iK)^2*d + B*i*K*D^2)*L$. As for MetaGR, since it transforms each interaction of an item in the sequence into two tokens, one representing the information of the item and the other representing the information of the action type, the input of the length of the token sequence to the model becomes $2N$. 
The computational complexity is $2*O(4*B*2N*D^2 + 2*B*H*(2N)^2*d + B*2N*D^2) * L$. In practical application scenarios, the sequence length $N$ is often much greater than $D$.
With this approximation, we can obtain that the ratio of computational complexity between SFGR and MetaGR is about $O(\frac{N}{12*K})$.

Based on our computational analysis, SFGR significantly reduces inference-stage computational demands to just one-fourth of MetaGR's approach, making it particularly valuable for industrial applications with strict online latency constraints. However, during the training stage, the computational ratio between SFGR and MetaGR is approximately $N:12K$. In real industrial recommendation scenarios, $N$ represents the total length of the user behavior sequence over extended time periods, often reaching thousands or even tens of thousands of interactions. Meanwhile, $K$, which represents the average number of items that a user interacts with in a session, is typically two to three orders of magnitude lower than $N$. From this perspective, the computational complexity introduced by SFGR's user sequence sample splitting during training is prohibitively high, increasing computational costs by 10 to 100 times and severely limiting its practical application in real industrial settings.

\subsection{Efficient Generative Ranking via dual-Flow Training}
\label{sec:dual-flow}

\begin{figure*}[t]
    \centering
    \includegraphics[width=\textwidth]{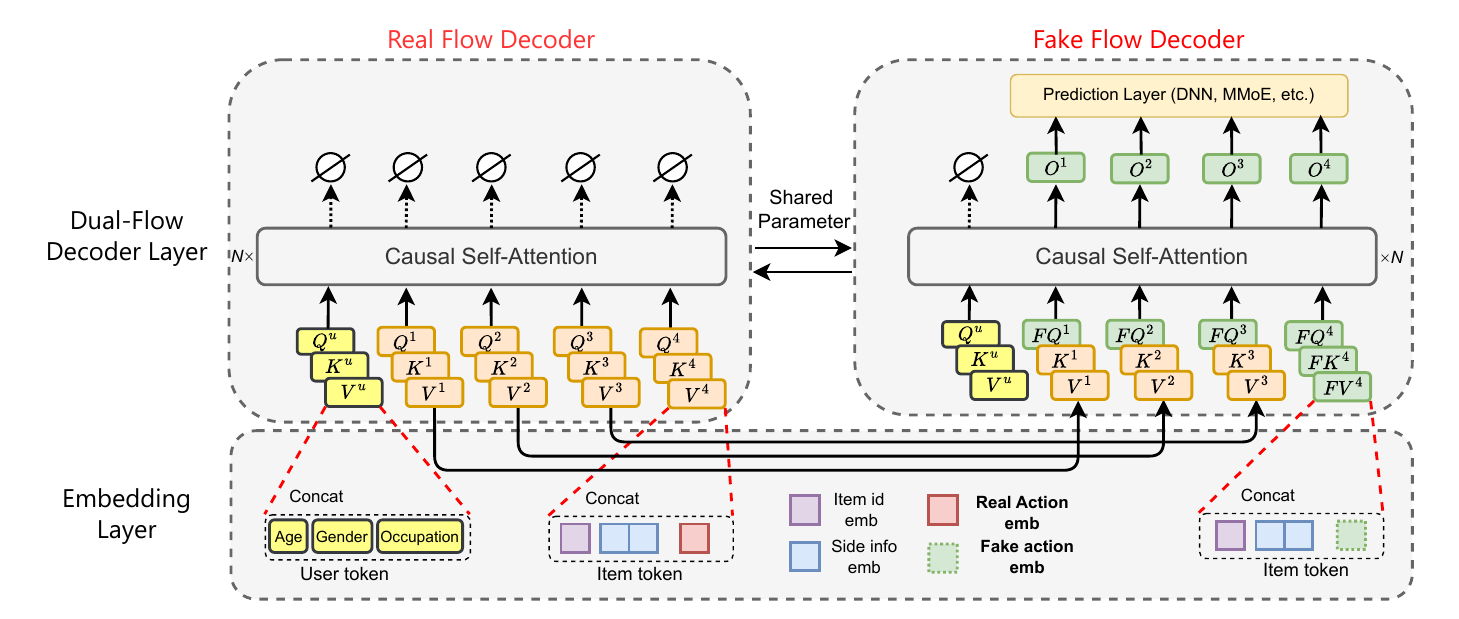}
    \caption{Model Architecture of Dual-Flow Generative Ranking Network.}
    \label{fig:flowchart}
\end{figure*}

\subsubsection{Model Architecture}
In order to ensure efficient computational performance during the inference stage while significantly reducing computational complexity during the training stage, we propose a dual-flow generative ranking network~(DFGR) in Figure~\ref{fig:flowchart}. As described in Section 3.1, under the generate ranking framework, each user's sequence constitutes a sample in which every token in the sequence is aggregated from multiple slot features representing item/user interaction information at each position. Notably, under the training-inference efficient dual-flow structure, the original behavior sequence is duplicated into two input vector sequences based on whether the action type in items uses real values, where one is called the fake flow and the other is called the real flow. In the real flow, tokens at positions corresponding to item elements incorporate real action-type slot information (\textit{e.g.} view/click/pay), which can be regarded as the target information for the model to learn. Conversely, in the fake flow, the action type uses a fake identifier serving as a placeholder that contains no actual action-type information about the item at that moment. 

Both flows share the network weights of their self-attention computation layers. During forward propagation, the real flow computes multi-head QKV attention following standard decoder-only self-attention mechanisms, though its output tokens are completely masked during loss calculation. For the fake flow computation, taking self-attention calculation of position $4$ as an example which shown in the figure ~\ref{fig:flowchart}~. Firstly, it derives $FQ^{4},FK^{4},FV^{4}$ through standard weight projection matrices $W_Q,W_K,W_V$, and in the decoder-only architecture, its attention computation relies on KV sequence information from both the current position and preceding tokens. For KV activation states at positions preceding $4$ in the sequence, we utilize $[K^1,K^2,K^3]$ and $[V^1,V^2,V^3]$ from the corresponding layer in the real flow containing actual action types, while for KV activation states at the current position, we employ
$FK^4$ and $FV^4$. This approach simultaneously prevents label information leakage during self-attention propagation while enabling each token to comprehensively capture all preceding contextual information, including action-type details. When the fake flow propagates through N decoder-only Transformer layers, The final output embedding can be connected to various prediction heads such as DNN, MMOE \cite{ma2018modeling}, or other architectures for pCTR/pCTCVR prediction. During training, the loss computation occurs only for tokens positioned in the fake flow. From a holistic perspective, each token in the sequence serves dual roles by acting as a labeled sample for loss computation at its corresponding timestep when processed in the fake flow, while functioning as contextual features for subsequent tokens when present in the real flow.

The aforementioned computational process of DFGR can be applied to any decoder-only Transformer variant framework. In practical implementation, one may adopt the standard decoder-only transformer architecture, which consists of self-attention with softmax normalization followed by feedforward networks. Alternatively, one may employ the GAU module\cite{hua2022transformer} which reduces model complexity by removing FFN layers or adopt Meta's HSTU architecture. Given HSTU's lower computational complexity and proven effectiveness in industrial recommendation scenarios, the following formulations are based on this structure.

{\small
\begin{equation}
U^{r},\ V^{r},\ Q^{r},\ K^{r}=\mathrm{Split}\left(\sigma_1(f_1(X^{r}))\right) \tag{1}
\end{equation}

\vspace{-3ex}
\begin{equation}
\begin{split}
A^{r}·V^{r} = \sigma_2\left(Q^{r}·{K^{r}}^T\odot Mask + rab^{p,t}\right)·V^{r}
\end{split} \tag{2}
\end{equation}

\vspace{-3ex}
\begin{equation}
Y^{r} = f_2\left(\mathrm{Norm}\left(A^{r}V^{r}\right) \odot U^{r}\right) \tag{3}
\end{equation}
}

\vspace{3ex}
Equations (1) to (3) describe the forward computation process of the real flow which is the same as that in MetaGR.

{\small
\begin{equation}
\begin{split}
U^{f},\ V^{f},\ Q^{f},\ K^{f} = \mathrm{Split}\left(\sigma_1(f_1(X^{f}))\right)
\end{split} \tag{4}
\end{equation}

\vspace{-2ex}
\begin{equation}
\begin{split}
A^fV^r = \sigma_2(Q^f{K^r}^T\odot Mask + rab^{p,t})V^r
\end{split}
\tag{5}
\end{equation}

\vspace{-2ex}
\begin{equation}
\begin{split}
A^{f}_{i}V^{f}_{i} &= \sigma_2(Q^{f}_{i}K{^{f}_{i}}^T + rab^{p,t}) V^{f}_{i}
\end{split} \tag{6}
\end{equation}

\vspace{-2ex}
\begin{equation}
\begin{split}
Y^f = f_2\Big(&\mathrm{Norm}(A^{f}_{i}V^{f}_{i} + A^{f}V^{r}\odot(\mathbf{1}_t-\mathbb{I}_t)) \odot U^{f}\Big)
\end{split}
\tag{7}
\end{equation}
}
Equations (4) to (7) outline the forward computation process of the fake flow.

Where $f_{i}(x)$ denotes an MLP; we use one linear layer, $f_{i}(x)=w_ix+b_i$ for $f_1$ and $f_2$ to reduce compute complexity and further batches computations for queries Q, keys K, values V, and gating weights U with a fused kernel; $\sigma_1$ and $\sigma_2$ denote nonlinearity, for both of which we use SiLU\cite{elfwing2018sigmoid}; Norm is layer norm; and $rab^{p,t}$ denotes relative attention bias\cite{raffel2020exploring} that incorporates positional (p) and temporal  (t) information; Mask denotes mask. $X^f$ represents the input of fake flow, while $X^r$ represents the input of real flow. Given a total input sequence length of t, the notation without subscripts $X^f$ denotes elements $1$ through $t$ in the fake flow. The symbol $\mathbf{1}_t$ denotes the all-ones matrix, while $\mathbb{I}_t$ denotes the identity matrix.

\paratitle{Cross-Triangle Matrix For Intra-Session Information Masking.}
\begin{figure}[t]
    \centering
    \includegraphics[width=0.7\linewidth]{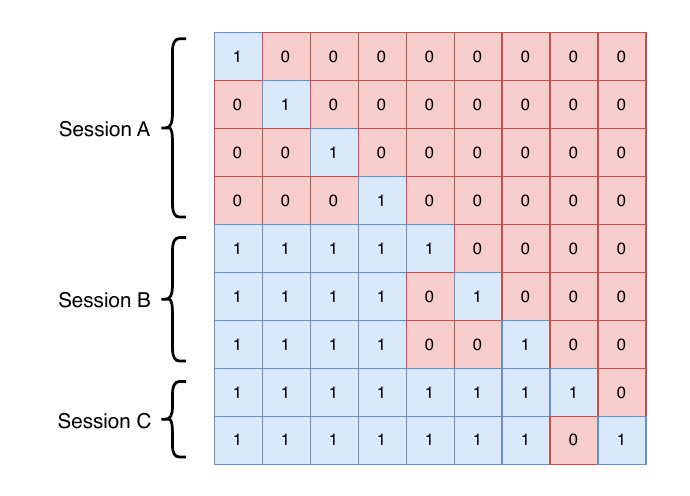}
    \caption{Session-Aware Cross Triangle Mask.}
    \label{fig:enter-matrix}
\end{figure}
Unlike NLP tasks, though behavior sequences in DFGR are chronologically ordered, real-world user interactions consist of independent sessions, each initiated by the same recommendation request, with all items in a session being scored simultaneously by the ranking system. Thus, while items within a session may have varying timestamps (reflecting view/click/purchase times),  they should remain mutually unaware during parallel pointwise scoring to avoid label leakage. In order to achieve this, we modify the standard lower triangular mask by zeroing out attention between items sharing the same session id. For example, consider a user sequence of length 9, which is divided into 3 sessions containing 4, 3, and 2 items respectively. We construct the session-aware cross-triangle mask matrix as shown in Figure ~\ref{fig:enter-matrix}~. The zero red entries in the matrix indicate the modifications required for generative recommendation scenarios, which effectively prevent mutual awareness between items within the same session.
When making predictions, the same strategy can be employed as the inference-efficient single-flow structure described in Section 3.2, which will not be repeated here.

\subsubsection{Comparative Analysis of Computational Complexity}
We follow Section~\ref{sec:single-flow} to conduct a comparative analysis of computational complexity between our DFGR and MetaGR.

\paratitle{Inference.}
Since DFGR adopts the same structure and data organization method as SFGR during inference, it also achieves a 4x speedup over the MetaGR model.

\paratitle{Training.} 
Recall that in Section~\ref{sec:single-flow}, the computational complexity of training for MetaGR is $2*O(4*B*2N*D^2 + 2*B*H*(2N)^2*d + B*2N*D^2) * L$.
For our DFGR, the computational complexity is $2*2*O(4*B*N*D^2 + 2*B*H*N^2*d + B*N*D^2)*L$. 
In practical application scenarios, the sequence length $N$ is often much greater than $D$.
With this approximation, we can obtain that the ratio of computational complexity between our DFGR and MetaGR is about 2.

Under the DFGR's approach, the sequences in the training process are duplicated into real flow and fake flow, causing the computational load to double compared to a single flow. The MetaGR's approach splits each interaction into interleaved sequences composed of items and action-types, which effectively doubles the input sequence length while processing the same user history. Since self-attention operations, which constitute the main structure of the model, scale quadratically with sequence length, the DFGR approach saves half of the computational resources during training compared to MetaGR. In summary, compared to the MetaGR's approach, DFGR features a lower theoretical computational complexity in both the training stage and inference stage, positioning it as the next-generation generative ranking architecture.

\section{Experiment}
Regarding the DFGR approach, here are some questions we need to address:

(1) How does DFGR perform in offline evaluations compared to other SOTAs and MetaGR?

(2) What are the effects of key components and different implementations in the proposed method? (\textit{e.g.}, MoA multi-head/optimizer scheduling strategy/full-domain data)

(3) How to determine the optimal parameter allocation strategy in real-world industrial deployment scenarios?

(4) Can the DFGR approach observe the scaling law phenomenon?

\subsection{Experimental Setup}

\paratitle{Datesets.}
We selected three datasets for evaluation: RecFlow\footnote{https://github.com/RecFlow-ICLR/RecFlow}, KuaiSAR\footnote{https://kuaisar.github.io/}, and Trec, an industrial dataset from a major e-commerce platform in China. Both RecFlow and KuaiSAR are public Kuaishou datasets containing user log data from their video platform. These datasets exhibit richer user behavior patterns compared to conventional benchmarks, which makes them particularly suitable for evaluating the performance of the DFGR model. TRec uses real industrial-grade user behavior data from a major e-commerce platform in China, which includes approximately 30 million users and covers behaviors over the past year, with sequence lengths truncated to 4k. In addition to recommendation channels, we also incorporated user behavior data from search and other full-domain sources. During the evaluation, to prevent data leakage, data from the time range [1, T] are used as a training set, while data from time T+1 are used as a validation set.

\paratitle{Compared Baselines.}
To demonstrate the effectiveness of the DFGR model, we selected the following SOTA algorithms for comparison.

\textbullet~\textbf{DeepFM}\cite{guo2017deepfm}, consists of DNN and FM components. FM extracts low-dimensional cross features, while DNN extracts high-dimensional cross features.

\textbullet~\textbf{DIN}\cite{zhou2018deep}, applies the target attention structure to the user's historical sequence to obtain dynamic weights of historical behaviors to capture user interests.

\textbullet~\textbf{DIEN}\cite{zhou2019deep}, combines GRU and target attention, simultaneously capturing the evolution of user interests and the dynamic weights of historical behaviors.

\textbullet~\textbf{DCN}\cite{shan2016deep}, includes DNN and cross-components. Cross explicitly extracts higher-order cross features, while DNN implicitly extracts high-dimensional cross features. 

\textbullet~\textbf{SASRec}\cite{kang2018self}, is the first and most classic transformer-based sequential recommendation model with unidirectional causal self-attention.

\textbullet~\textbf{BERT4Rec}\cite{sun2019bert4rec}, employs deep bidirectional self-attention to model user behavior sequences and serves as a representative sequential recommendation model.

\textbullet~\textbf{MetaGR}\cite{zhai2024actions}, is a state-of-the-art generative recommendation model proposed by Meta, where items and actions are alternately organized to form a sequence of user behaviors, and the main body of the model consists of multiple layers of the HSTU structure. 

\textbullet~\textbf{DLRM}, as the currently deployed online baseline model for homepage recommendations on the e-commerce platform, is built on a state-of-the-art architecture incorporating DIN, SIM, and MMoE components, with extensive feature engineering.

\paratitle{Experimental Setting.}
As previously described, we conducted experiments using the three datasets. Specifically, data within the range of [1, T] was used as a training dataset, while data at T + 1 was used as the validation dataset. Regarding the evaluation metrics, we chose the Area Under Curve (AUC) and Group Area Under Curve (G-AUC), which are widely adopted in the academic community. Both AUC and G-AUC are numerical values ranging from 0 to 1, and a larger value of either metric indicates a stronger ranking ability of the model. 

For all compared SOTAs, we employed a substantial number of user and item features. In the cases of DIN and DIEN, users' historical behavior sequences were additionally incorporated into the model architectures alongside the aforementioned features. DFGR aims to achieve optimal implementation results by training on 4 nodes with 32 A100 GPUs, a batch size of 256, a network depth of 8 layers, and a learning rate of 5e-4, using full-domain data and annealing introduced in the final stages. To ensure a fair comparison, MetaGR is trained in the same configuration as DFGR. The DLRM SOTA utilizes the e-commerce platform's current recommendation baseline model, incorporating approximately hundreds of manual feature engineering features with a structure of DNN + MMOE + SIM for user behavior sequences, which were produced from full-domain data.

\subsection{Overall Performance (RQ1)}

 \begin{table}[t]
  \caption{Overall performance comparison on public datasets. The best and runner-up results are bold and underlined.}
  \label{tab:experimental_results}
\begin{tabular}{lcc|cc}
    \toprule
         \multirow{2}{*}{Method}&  \multicolumn{2}{c|}{RecFlow} &  \multicolumn{2}{c}{KuaiSAR}\\ \cline{2-5}
 & AUC & G-AUC & AUC & G-AUC \\ 
    \midrule
         SASRec & 0.6923 & 0.6815 & 0.6213 & 0.5987 \\ 
         BERT4Rec & 0.6933 & 0.6829 & 0.6226 & 0.5992 \\ 
         DeepFM & 0.6968 & 0.6830 & 0.6308 & 0.6035 \\ 
         DIN & 0.6942 & 0.6860 & 0.6238 & 0.6063 \\  
         DIEN & 0.6963 & 0.6865 & 0.6293 & 0.6065 \\ 
         DCN & 0.6986 & 0.6877 & 0.6379 & 0.6070 \\
         MetaGR  &  \underline{0.6990} &  \underline{0.6884} &  \underline{0.6395} &  \underline{0.6072} \\ 
    \midrule
 {DFGR}& \textbf{0.7012} & \textbf{0.6892} & \textbf{0.6472} & \textbf{0.6081}\\ 
 \bottomrule
  \end{tabular}
\end{table}

\begin{table}[t]
  \caption{Overall performance comparison on industrial dataset TRec.}
  \label{tab:ICL experimental_results_industry}
  \begin{tabular}{l|cc}
    \toprule
    Method & AUC & G-AUC \\
    \midrule
    DLRM & 0.8703 & 0.7220 \\
    MetaGR & 0.8689 & 0.7185\\
    \midrule
    DFGR (rec only) & 0.8721 & 0.7227\\ 
    DFGR & \textbf{0.8755} & \textbf{0.7288} \\
    \bottomrule
  \end{tabular}
\end{table}

It is well-established that in recommendation scenarios, a 0.1\% improvement in AUC and G-AUC represents a significant enhancement that contributes substantial commercial value. As shown in Table \ref{tab:experimental_results}, we compared all models except DLRM on the RecFlow and KuaiSAR datasets. Table \ref{tab:ICL experimental_results_industry} demonstrates the comparison between DLRM, MetaGR, DFGR Single-Channel, and DFGR on the dataset TRec.

First, DFGR consistently outperforms other models in all datasets. In RecFlow and KuaiSAR datasets, generative models demonstrate superior performance compared to traditional DNN models, highlighting the strong modeling capabilities of the generative paradigm. On the TRec dataset, DFGR surpasses DLRM, which in turn outperforms MetaGR. This observation reveals that, while improperly designed generative paradigms might be outperformed by traditional DNN models that integrate various advantages and extensive feature engineering, well-designed generative approaches can break through the performance ceiling of conventional DNN models.

Second, DFGR achieves improvements of 0.31\%-1.2\% AUC and 0.11\%-1.4\% G-AUC over MetaGR in the three datasets, demonstrating the advantages of its dual-flow paradigm. In MetaGR's architecture, heterogeneous data of items and action-type are mapped into the same vector space, which substantially compromises model training effectiveness. In contrast, DFGR's dual-flow paradigm successfully circumvents this limitation inherent to MetaGR's approach, thereby achieving greater performance gains.

\subsection{Effects Of Key Components  And Strategys (RQ2)}

\subsubsection{Effective of full domain sequence data}

We conducted experiments to investigate the effects of the full-domain data. Comparative studies were performed using (1) RecOnly-channel data, user behavior sequences of the target scenario that refers to the homepage of the e-commerce platform, and (2) full-domain data, multi-scenario data including Rec channel and other channel in APP. To ensure experimental control, both models maintained identical configurations except for data sources:
(1) user behavior sequences were truncated to a maximum length of 4K,
(2) identical model architecture and training hyperparameters were employed, and
(3) the DNN output layers were partitioned into channels in the full domain model to preserve the consistency of the value distribution while maintaining equivalent activation volumes.

As shown in Table \ref{tab:ICL experimental_results_industry}, the full-domain model(DFGR) demonstrated superior performance with a 0.57\% AUC improvement and 0.84\% G-AUC improvement, compared to the single-channel model(DFGR(rec only)). These results indicate that DFGR effectively captures multi-dimensional user interests in full domains, thereby improving model effectiveness through cross-scenario interest representation.

\begin{figure}[t]
    \centering
    \includegraphics[width=1.0\linewidth]{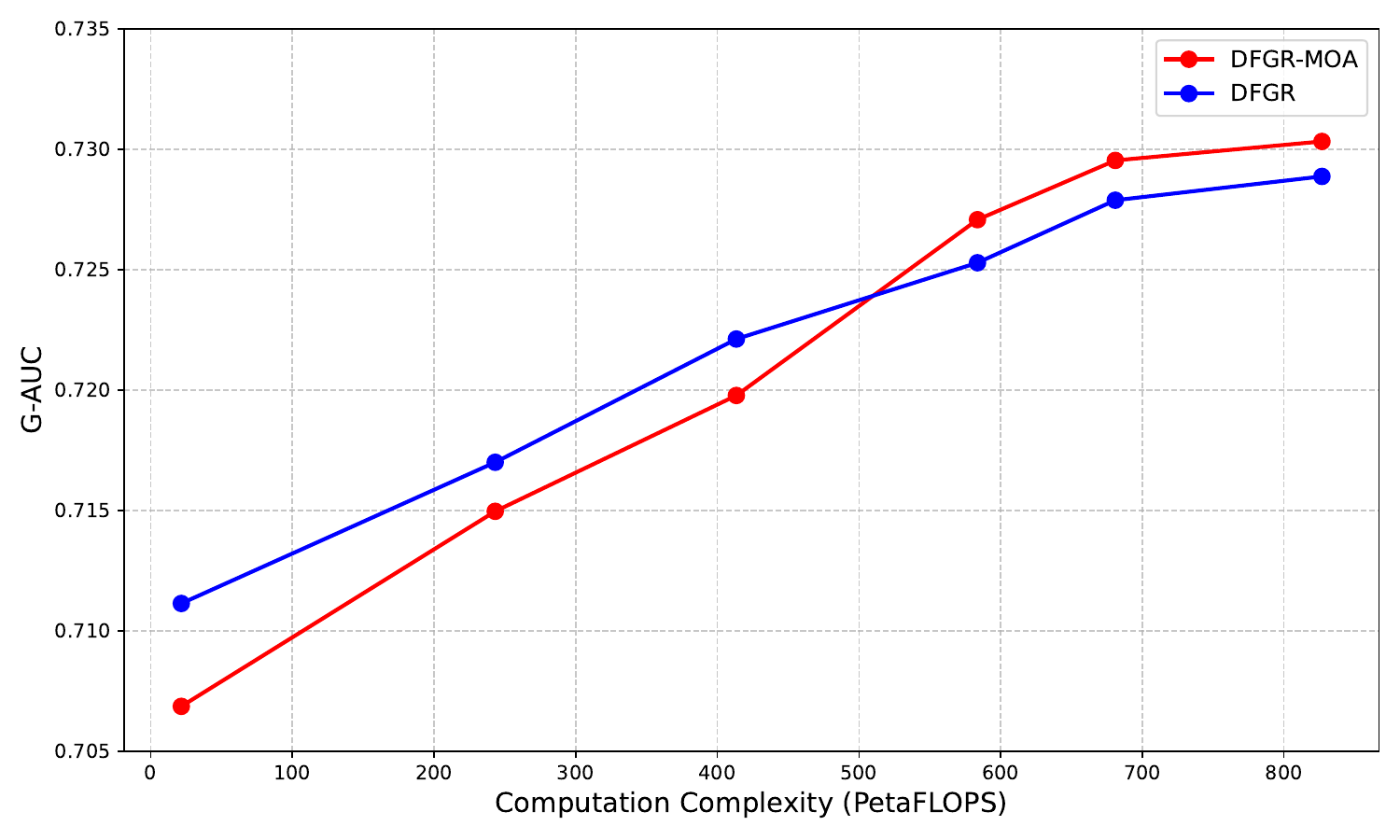}
    \caption{DFGR vs DFGR-MOA Performance.}
    \label{fig:moa_vs_dense_performance}
\end{figure}

\subsubsection{Effective of Mixture-of-Attention(MOA) in full domain sequence data}

As demonstrated in Section 4.5.1, the model uses full-domain data as input. To better leverage the distinct characteristics of different data sources in model operations, we implement a Mixture-of-Attention (MOA) \cite{zhang2022mixture} structure for computing multi-head self-attention, which we designate as DFGR-MOA. The specific implementation of the MOA structure comprises two distinct components: one portion of the heads are shared across all channels, while the other portion of heads remain exclusive to individual channels. Each token's representation is formed by concatenating outputs from both shared heads and channel-specific heads, ensuring balanced preservation of both commonality and individuality across different channels.

For fair experimental comparison, DFGR and DFGR-MOA maintain identical activated parameters and computational costs. Notably, under equivalent activated parameter conditions, DFGR-MOA engages more trainable parameters than DFGR. The experimental results (Figure \ref{fig:moa_vs_dense_performance}) reveal that DFGR-MOA initially underperforms DFGR before 50K training steps due to its larger parameter space requiring more training data, but consistently outperforms DFGR after 50K steps, with an improvement (G-AUC 0.28\%).

\begin{figure}[t]
    \centering
    \includegraphics[width=1.0\linewidth]{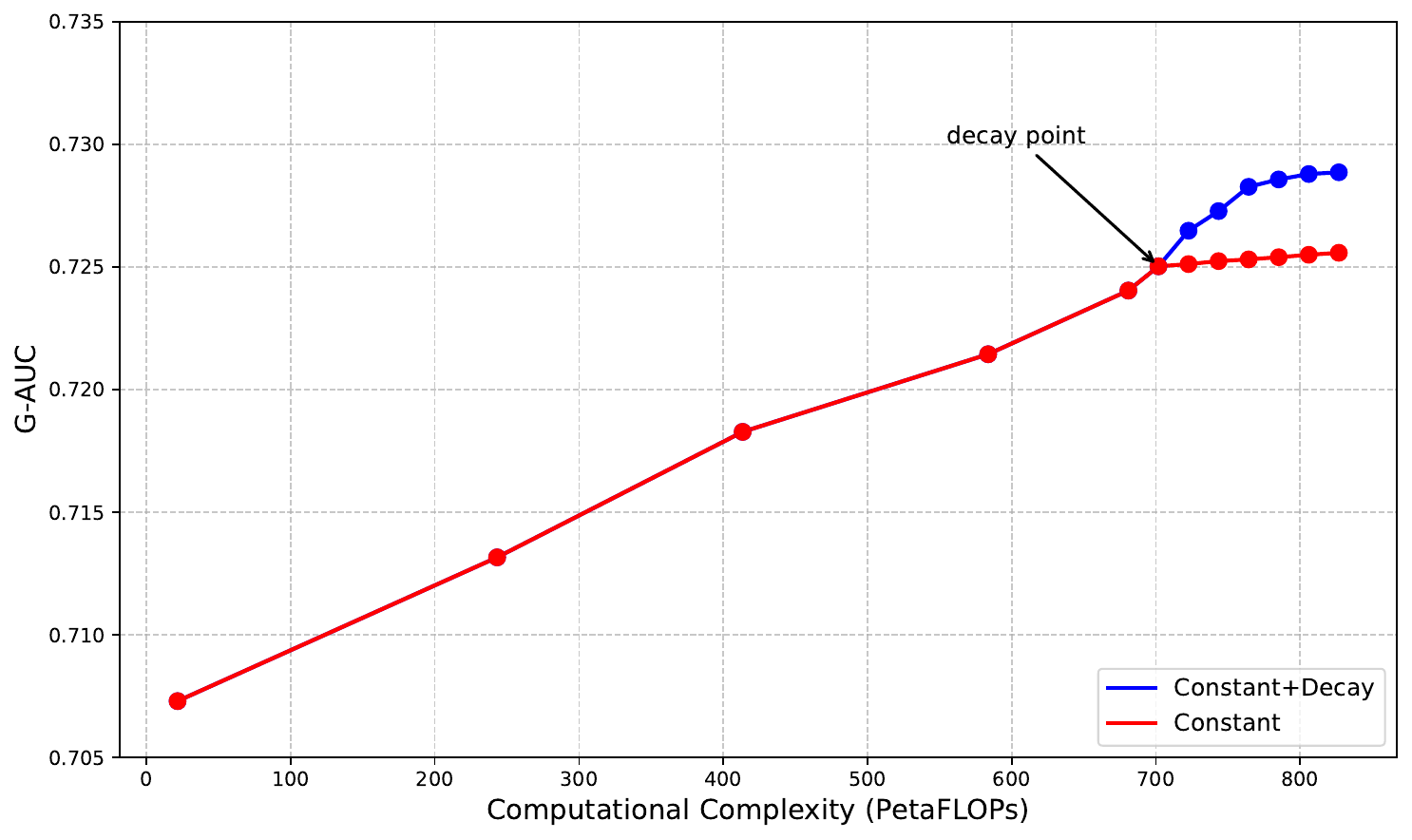}
    \caption{Constant vs Decay Performance.}
    \label{fig:constant_vs_decay_performance}
\end{figure}

\subsubsection{Effective of learning rate scheduling strategy}

Inspired by the WSD strategy~\cite{hu2024wsd} prevalent in the domain of large language models, we propose a modified training regimen that eliminates the warmup phase. This adjustment is motivated by empirical observations that the conventional warmup phase originally designed for ultra-deep LLM architectures demonstrates diminishing returns when applied to models with shallower configurations. Consequently, our approach partitions the training process into two distinct phases: 
(1) constant learning rate, where a fixed rate of 5e-4 accelerates model convergence, and
(2) decay learning rate, where the learning rate starts at 5e-4 and decays at a rate of 5e-6 per 1,000 steps.
To validate this approach, we conducted comparative experiments on the trained model:
(1) Control Group, continued training with the original constant learning rate, and
(2) Experimental Group, switched to the decay learning rate strategy.

As illustrated in Figure \ref{fig:constant_vs_decay_performance}, the experimental results exhibit remarkable similarity to the phenomena observed in the LLM training. The final annealing phase, characterized by rapid reduction in learning rate, effectively drives the model to converge to a local optimum, resulting in substantial improvements (G-AUC 0.42\%) in all evaluation metrics. This empirical evidence suggests that the decline in the strategic learning rate can enhance the efficiency of model optimization without requiring additional computational resources.

\begin{figure}[t]
    \centering
    \includegraphics[width=1.0\linewidth]{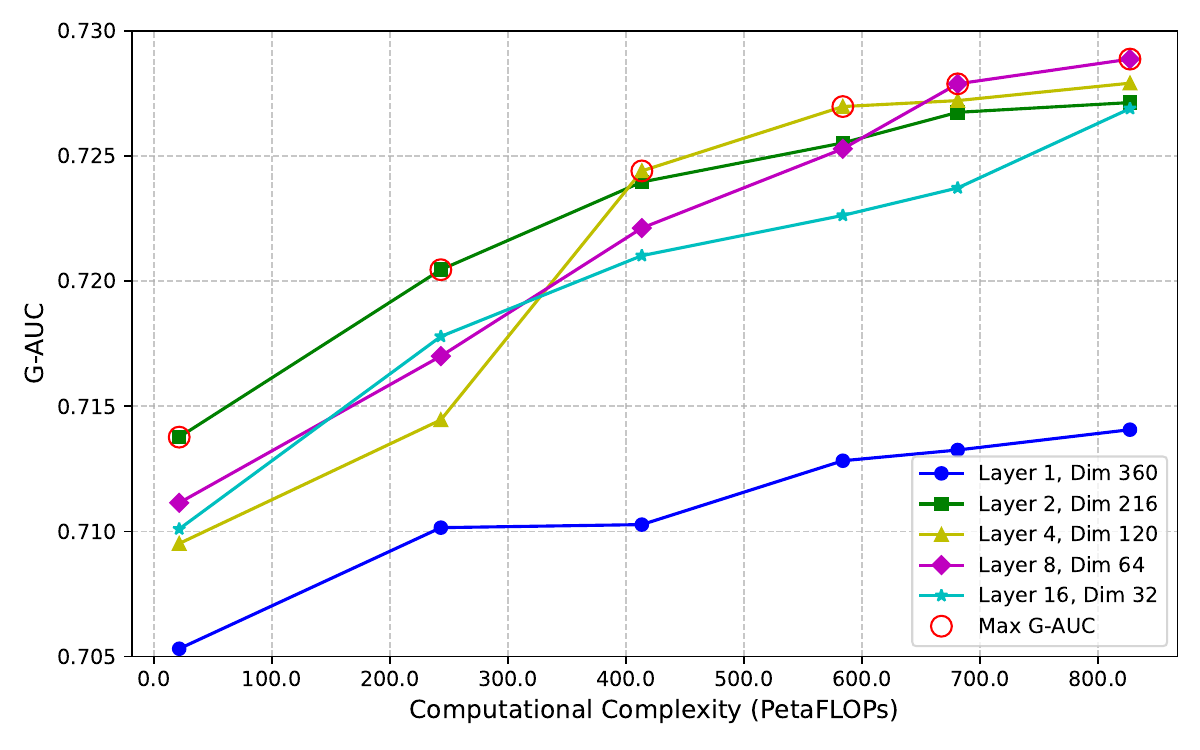}
    \caption{Performance Comparison of Different Model Parameter Settings.}
    \label{fig:Performance_Comparison_of_Different_Model_Parameter_Settings}
\end{figure}

\subsection{Optimal parameter allocation strategy under computational constraints (RQ3)}
\label{sec:potimal parameter}
To investigate optimal model parameter allocation strategies under computational constraints, we conducted a systematic study designing five experiments with equivalent computational budgets. Each experiment employed distinct architectural configurations, varying network depth from 1 to 16 layers and width from 32 to 360 units, while maintaining consistent training hyperparameters across all experiments.

As illustrated in Figure \ref{fig:Performance_Comparison_of_Different_Model_Parameter_Settings}, the shallowest configuration (1-layer) demonstrated significantly inferior performance in all training stages, primarily due to its limited capacity to model implicit nonlinear interactions within user behavior sequences. Conversely, excessively deep models (16-layer, where 'excessive depth' is a scenario-dependent relative concept) exhibited suboptimal performance across training phases, attributable to optimization challenges in gradient backpropagation. However, the 16-layer configuration manifests a robust catch-up trajectory, suggesting its potential to emerge as a SOTA model when deployed in scenarios with substantially larger datasets. This observation aligns with the scaling law principles observed in large language models \cite{kaplan2020scaling,hoffmann2022training}, emphasizing the importance of appropriate depth-width ratio configuration. 

The remaining three configurations revealed an evolutionary pattern. As computational expenditure increased, optimal network depth progressively shifted from 2-layer to 4-layer architectures and ultimately to 8-layer architectures. This dynamic demonstrates that the optimal model structure evolves with varying computational constraints. Our experiments indicate a consistent trend toward deeper architectures under increasing computational complexity - deeper networks exhibit enhanced capability in extracting complex implicit interaction patterns from user behavior sequences. However, this advantage is contingent upon sufficient availability of training data, as deeper architectures require more extensive parameter learning. Therefore, under appropriate depth-to-width ratios, as both training dataset size and computational resources increase, deeper network architectures have progressively demonstrated superior performance under equivalent computational complexity, ultimately emerging as the optimal configuration choice.

\begin{figure}[t]
    \centering
    \includegraphics[width=1.0\linewidth]{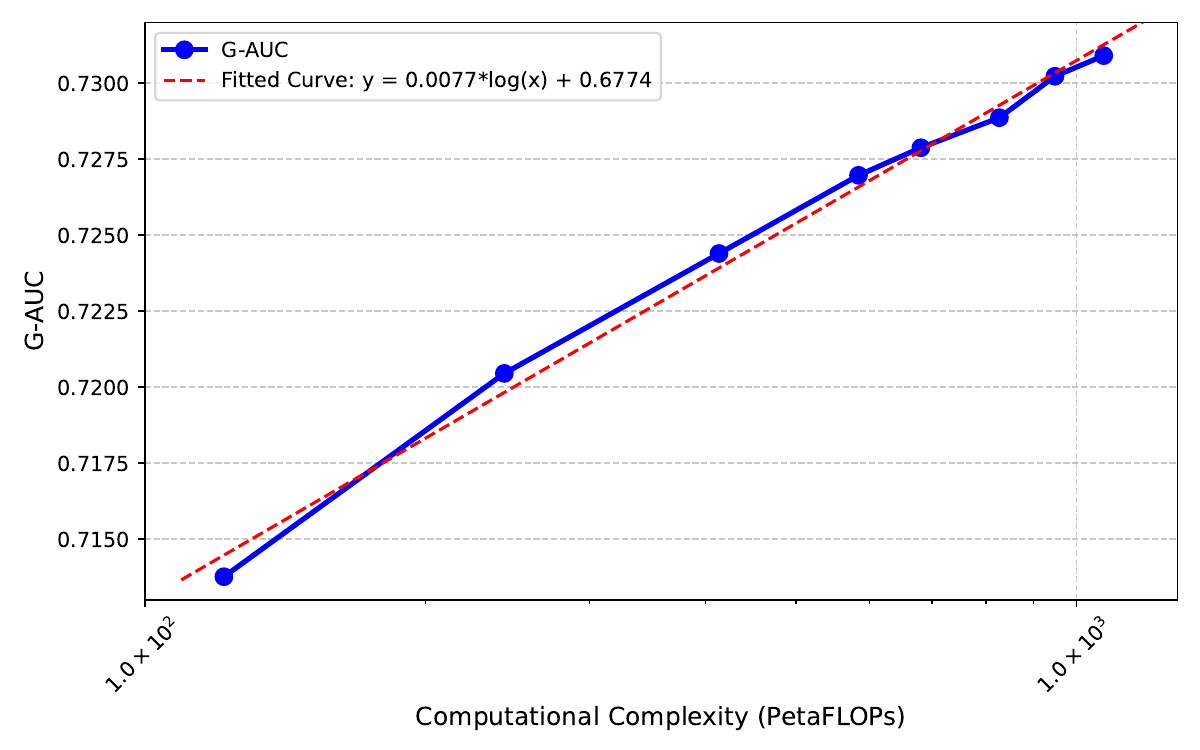}
    \caption{Scaling Laws with Computational Complexity.}
    \label{fig:Scaling_Law_Curve_of_DFGR}
\end{figure}

\subsection{Scaling Law in DFGR (RQ4)}
As demonstrated in Section \ref{sec:potimal parameter}, the parameter configurations of DFGR corresponding to optimal models under different computational complexity, together with their respective optimal G-AUC metrics, are summarized. We plot the computational complexity (represented on a logarithmic scale as the horizontal axis) against the corresponding optimal G-AUC values (vertical axis), as illustrated in Figure \ref{fig:Scaling_Law_Curve_of_DFGR}. The results reveal that in the current scenario, the DFGR model adheres to the scaling law principle, where the G-AUC metric exhibits a linear improvement trend as computational complexity increases following a power-law pattern.

\section{Conclusion}
In this paper, we introduce the Dual-Flow Generative Ranking Network (DFGR), a novel architecture that addresses fundamental challenges in recommendation systems. Using a dual-flow mechanism with innovative interaction methods between real and fake flows in the self-attention mechanism, our approach eliminates the need for extensive manual feature engineering while significantly improving both training and inference efficiency.

The experimental results across both open-source benchmarks and real industrial datasets demonstrate DFGR's superior performance compared to traditional Deep Learning Recommendation Models and other state-of-the-art approaches. Our model consistently outperforms the DLRM baselines, Meta's HSTU approach, and common architectures such as DIN, DCN, and DeepFM across key ranking metrics, validating its effectiveness in real-world recommendation scenarios.

Furthermore, our systematic investigation into optimal parameter allocation strategies under computational constraints provides valuable insight for practical deployment in industrial settings with strict latency requirements. The scaling law observations and learning rate scheduling strategies we explored offer a solid foundation for future research and development in this direction.

DFGR represents a significant step forward in generative ranking paradigms, offering a more efficient, effective, and scalable solution for modern recommendation systems. By reducing computational overhead while maintaining or improving performance, our approach bridges the gap between academic research and industrial application requirements. We believe that this work establishes a foundation for the next generation of state-of-the-art generative ranking paradigms, opening promising research directions for automated feature engineering recommendation approaches that effectively utilize raw behavioral data while maintaining computational efficiency.

\section*{GenAI Usage Disclosure}
In preparing this paper, we used GPT-4 and Deepseek-V3 to identify and correct grammatical errors, typos, and other writing mistakes in the initial draft. No AI tools were used for data analysis, experimentation, or the formulation of conclusions. We acknowledge the contributions of GPT-4 and Deepseek-V3 to enhance the writing process while maintaining full academic integrity.


\bibliographystyle{ACM-Reference-Format}
\bibliography{sample-base}


\end{document}